%% file: 0_Main.tex
\documentclass[journal,12pt]{IEEEtran}

\usepackage{cite}
\usepackage[usenames, dvipsnames]{color}
\usepackage{graphicx}
\usepackage[cmex10]{amsmath}
\usepackage{array}
\usepackage{fixltx2e}
\usepackage{dblfloatfix}
\usepackage{relsize}
\usepackage{amsmath,amssymb,amsthm}
\usepackage{dsfont}
\usepackage{amsfonts}
\usepackage{graphicx}
\usepackage{cite}
\usepackage{pgf}
\usepackage{tikz}
\usepackage{rotating}
\usepackage{setspace}
\usepackage{enumerate}
\usepackage{pifont}
\usepackage{mathtools}
\usepackage{tikz}
\usepackage{hyperref}
\usepackage[font=small,labelfont=bf, tableposition=top]{caption}

\usetikzlibrary{chains,shapes.multipart}
\usetikzlibrary{shapes,calc,fit}
\usetikzlibrary{automata,positioning,patterns,decorations.pathreplacing}
\usetikzlibrary{arrows,automata,patterns,decorations.pathreplacing,positioning,calc,chains,shapes.multipart,shapes,fit}
\usepackage{algorithm}
\usepackage{algorithmic}
\usepackage{bm}
\usepackage{url}
\usepackage{subcaption}
\usepackage{mwe}

\theoremstyle{definition}
\newtheorem{dfn}{Definition}
\newtheorem{app}{Approximation}

\newtheorem{cor}{Corollary}
\newtheorem{exm}{Example}

\newtheorem{theorem}{Theorem}

\newcommand{\um}[1]{${U}_{#1}$}
\newcommand{\uu}[1]{{U}_{#1}}
\newcommand{\cm}[1]{$c_{#1}$}
\newcommand{\cc}[1]{c_{#1}}
\newcommand{\cbar}[1]{\bar{c}_{#1}}
\newcommand{\pma}[1]{$\mathbf{p}_{#1}$}
\newcommand{\pp}[1]{\mathbf{p}_{#1}}
\newcommand{\nm}[1]{$N_{#1}(t)$}
\newcommand{\nn}[1]{N_{#1}(t)}
\newcommand{\rate}[1]{R_{#1}}
\newcommand{\dd}[1]{d_{#1}(t)}
\newcommand{\dm}[1]{$d_{#1}(t)$}
\newcommand{\hh}{\mathcal{H}}
\newcommand{\LL}{\mathcal{L}}
\newcommand{\tick}{\mbox{\ding{52}}}
\newcommand{\cross}{\mbox{\ding{56}}}

\begin{document}
	
\title{Multirate Packet Delivery In Heterogeneous Broadcast Networks}

\author{Mohammad A. Zarrabian, Foroogh S. Tabataba, Sina Molavipour and Parastoo Sadeghi

\thanks{Copyright (c) 2015 IEEE. Personal use of this material is permitted. However, permission to use this material for any other purposes must be obtained from the IEEE by sending a request to pubs-permissions@ieee.org.
	
Mohammad A. Zarrabian and Foroogh S. Tabataba are with the Department of Electrical and Computer Engineering, Isfahan University of Technology, Isfahan, Iran 84156-83111. Emails: ma.zarrabian@ec.iut.ac.ir and fstabataba@cc.iut.ac.ir.
		
Sina Molavipour is with the Department of Electrical Engineering, KTH Royal Institute of Technology, Stockholm, Sweden. Email: Sinmo@kth.se.

Parastoo Sadeghi is with the Research School of Electrical, Energy and Materials Engineering, College of Engineering and Computer Science, The Australian National University, Canberra, Australia. Email: parastoo.sadeghi@anu.edu.au.}}

\maketitle

\begin{abstract}
	In this paper, we study the problem of multirate packet delivery in heterogeneous packet erasure broadcast networks. The technical challenge is to enable users receive packets at different rates, as dictated by the quality of their individual channel. We present a new analytical framework for characterizing the delivery rate and delivery delay performance of a previously proposed non-block-based network coding scheme in the literature. This scheme was studied in homogeneous network settings. We show for the first time, via new theoretical analysis and simulations that it can actually achieve multirate packet delivery. Using acknowledgments from each user, we show that the user with the highest link capacity achieves the maximum possible throughput.  Also, a non-zero packet delivery rate is possible for other users, and the delivery rate depends on the difference between the packet arrival rate at the sender and the link capacity of each user. The accuracy of our analytical framework is confirmed by comparing the results with simulations for different settings of packet arrival rate at the sender and link capacities.
\end{abstract}


\input{1_Introduction.tex}
\input{2_System-model.tex}
\input{3_Delivery_Rate_Analysis.tex}
\input{4_Delay_Analysis.tex}
\input{5_Simulation_Results.tex}
\input{6_Conclusions.tex}

\section{Acknowledgement}
The authors would like to acknowledge helpful discussions with Prof. Muriel M{\'e}dard in early parts of this work.
\bibliographystyle{IEEEtran}
\bibliography{IEEEabrv,Refrence}
\vspace{-1cm}

\end{document}

%% file: 1_Introduction.tex
\section{Introduction} 
Nowadays with the pervasive development of  wireless communication networks, the real-time applications such as broadcast multimedia and video streaming with high quality are in high demand \cite{2016SVCstreamloading,2017multidatanetwork,2018cashvehicular,2018trafficoffload}.
In wireless broadcast  streaming, an identical message is intended to be delivered in the form of ordered data packets to each user. 
An appropriate model for such a system is a packet erasure model, where a single sender aims to deliver ordered data packets  to some users over independent wireless packet erasure channels. 
When all the links, connecting to all the users have identical erasure probabilities, the network is called homogeneous, otherwise it is called heterogeneous \cite{2005sundarammultirate}.
In both homogeneous and heterogeneous networks, packet erasure events occur independently among users. 
Therefore, at any given time, the packets already received and still wanted at each user vary. This makes the design of  transmission schemes challenging.

An efficient method to accommodate multiple users' demands, achieving high throughput and decreasing delay is network coding  \cite{2000ahlswede,2003linear,2003koetteralgebraic,2006horandom}.
This method has been studied in different types of networks such as multicast and unicast networks \cite{2006horandom,2006parkcodecast:block,2006efficientunicast,2016QianUnicast,2018Garridomulticast}, multiple access and relay networks \cite{2014NCmultipleaccess,2015NCmultipleaccessII,2012multipleaccessRELwireless,2015multipleaccesssRElay}  as it is used to exploit the broadcast property of wireless channels and also to combat the packet erasure problem in networks \cite{2006danacapacity,2008XOR,2008OnlineBroadcasting,2015MaoWireless,2017garridosparse,2018khabazian}.

Most of the network coding methods are block-based \cite{2008eryilmazdelay:block,2010Eryilmazrandom,2010sorourminimum:block,
2011capacity:block,2015Sorour}, where a block of  packets is considered and a linear combination of the packets is constructed as the transmitted packet. 
In some cases, the transmitted packet is encoded in such a way that it provides new information for the most possible number of users. It is called innovation guarantee property. However, since the decoding of such methods depends on the reception of  the whole block, it may cause long delay in real-time packet streaming. 
Besides, in heterogeneous networks, another challenge is to provide packets for each user with respect to the quality of its channel, which is known as multirate packet delivery \cite{2005sundarammultirate}.
In block-based codes, encoded blocks with the length of $n$ packets convey $k\leq n$ packets of information and the rate of encoding would be $k/n$. 
In heterogeneous networks, the users with the link capacity lower than the encoding rate (weak users) cannot decode the packets and if the sender decreases the encoding rate, users with the higher bandwidth will experience long delays and their delivery rates decrease to the rate of the weak ones. 
To achieve multirate packet delivery in block-based codes, the sender must change the encoding rate which is inefficient and hard to implement when the number of users increases \cite{2013Multirateseq}.

For the purpose of real-time broadcasting, a coding method is preferred that would allow intermediate decoding of the packets  prior to the reception of the whole block \cite{2004RToblivious,2004MartinianThesis,2007RatelessCodes}. 
In these applications, average per packet delay is more important and due to the necessity of applying packets in-order, the performance of the system can be mainly measured by  the delivery rate (which is proportional to the throughput) and delivery delay rather than the decoding rate and  decoding delay. A packet is said to be delivered to a user  when all the previous packets with lower indices in the user's buffer have been decoded.

In addition to block-based codes, there also exist non-block-based network codes \cite{2009barroseffective,2008sundararajanarq,2008Sundararajanthreeuser,2009sundararajanfeedback,2014Sadeghidynamic,futhesis,2011sundararajanTCP,2012wureliable}.
Similar to \cite{2004RToblivious,2007RatelessCodes}, the goal in papers \cite{2008sundararajanarq,2008Sundararajanthreeuser,2009sundararajanfeedback,2014Sadeghidynamic,futhesis} is to increase the chance of decoding the packets prior to receiving all the information sent from the sender. 
They use the users' feedback to determine which packets should be encoded together and transmitted. 
For the purpose of real-time streaming, an ARQ (automatic repeat request) online network coding has been introduced in \cite{2008sundararajanarq} that combines the benefits of ARQ and network coding for broadcast networks.
It achieves the maximum throughput of one hop multicast networks but suffers from large delay for weak users.
For the delay mitigation problem, some solutions have been proposed in \cite{2009barroseffective,2008Sundararajanthreeuser,2009sundararajanfeedback}.
In \cite{2008Sundararajanthreeuser}, a non-block-based algorithm for three users has been suggested  and then it  has been improved for any number of users in \cite{2009sundararajanfeedback}. 
However, both  \cite{2008Sundararajanthreeuser} and \cite{2009sundararajanfeedback} only considered homogeneous networks. 
The authors in \cite{2009sundararajanfeedback} conjectured that their approach is asymptotically optimal in the decoding delay and delivery delay in the limit when  the packet arrival rate at the sender approaches the capacity (or the load factor approaches one). 
A non-asymptotic analysis (with respect to the load factor) of the works in \cite{2008sundararajanarq} and \cite{2009sundararajanfeedback} has been done in \cite{2014Sadeghidynamic,futhesis,2012SadeghiDeliverydelayanalysis} for homogeneous networks. 
In \cite{2012SadeghiDeliverydelayanalysis}, the authors have shown that the coding scheme of \cite{2009sundararajanfeedback} is more practical than the one in \cite{2008sundararajanarq}, because it provides more opportunities to transmit uncoded packets, which results in better decoding for the users. 
Based on the observations of \cite{2012SadeghiDeliverydelayanalysis}, a dynamic rate adaptation scheme was proposed in \cite{2014Sadeghidynamic} to improve system throughput and delay. 

\subsection{Approach and contributions }
In this paper, we provide an in-depth modeling and analysis of multirate packet delivery of non-block-based network coding of \cite{2009sundararajanfeedback} in heterogeneous networks. 
In the following we summarize our contributions and highlight distinctions with earlier works.

\begin{itemize}
\item {We demonstrate that the coding scheme proposed in \cite{2009sundararajanfeedback}, can indeed achieve multirate packet delivery in heterogeneous broadcast networks. 
Ensuring innovation guarantee property for all users and instantaneous delivery for some, this coding scheme achieves maximum possible throughput for the  user with the highest link capacity and a non-zero delivery rate for the others. 
The system model and the coding scheme is presented in Section \ref{sec:System model}.}

\item{Then, an analysis of the delivery rate of the coding scheme is proposed. Although the analysis of non-block-based codes is a challenging problem, using a reasonable approximation, we develop a tractable model to estimate the delivery rate.  To validate the analysis, our results are compared with extensive simulations for the different settings of the packet arrival rate and the channel capacities. 
Due to the existence of a transmission queue, this coding scheme is not deterministic as the one in  \cite{futhesis} (which also uses a different non-block-based network coding scheme) and the demand of the users is restricted by the packet arrival rate at the sender. This difference underpins our analysis  and the model is completely different from the ones used in \cite{futhesis}. 
The analytical model  of the delivery rate is discussed in Section \ref{sec:Delivery Rate Analysis}.}

\item{Finally, we analyze the delivery delay of the system based on a different definition from the previous works. 
In the literature, the delivery delay of a packet have been considered as the time when the packet enters the transmission queue to  the time it is delivered. However, we define the delivery delay as the time between the first request of a packet and its delivery. 
We believe this new definition is more suitable for the heterogeneous case and better characterizes the delay of the users. Consequently, it results in a simple closed-form delivery delay. To estimate and calculate the delivery delay, our delivery rate model is used  for different cases, which is shown  the consistency of our assumptions and approximation. Furthermore, the accuracy of the delivery rate and delivery delay analysis is confirmed by comparing the results with simulations. The delivery delay analysis is described in Section \ref{sec:Delay Analysis},  the simulation results in Section \ref{sec:Simulation Results}  and the paper is concluded in Section \ref{sec:Conclusions}.}	
\end{itemize}

%% file: 2_System-model.tex
\section{System Model} \label{sec:System model}
A single transmitter aims to broadcast a set of packets $\pp{1},\pp{2},\ldots,\pp{n}$ ($n$  arbitrary large) to  $\nu$ users ${\uu{i}},$ ($1\leq{i}\leq{\nu}$) via  heterogeneous broadcast  packet erasure channels. Here, a time-slotted scheme ($t=1,2,\ldots$) is assumed in which the sender uses linear network coding to construct the encoded packet for transmitting one coded packet in each time slot. Packets enter an infinite-length buffer, or the transmission queue at the sender according to a Bernoulli process of rate $\lambda$. Assuming independent channels between the transmitter and the users, each packet is correctly received by a user \um{i}  with a probability \cm{i} which is called the \textit{channel capacity}, i.e., packets are erased in each channel independently with the probability of $\cbar{i}=1-c_i$.\footnote{For simplicity, we use the notation $\bar{x}$ for $1-x$.}
Due to the heterogeneous property of the channels, the capacities are unique. Hence, without loss of generality, it is possible to assume that $\cc{1}>\cc{2}>\ldots>\cc{\nu}$, which is shown by the vector $\boldsymbol{c}=[\cc{1},\cc{2},\ldots,\cc{\nu}]$. Here, we indicate the strength of a user by its link capacity. We refer to $U_1$ as  the strongest user (i.e., with the highest link capacity).
The purpose of the system is to achieve multirate in-order packet delivery such that, more packets are delivered to the stronger users.

Each transmission is a linear combination of the packets along with a coefficient vector that determines the coefficient of each packet. The users store the received  packets and the coefficient vectors in their buffers to apply Gaussian elimination for decoding. The coefficients are chosen from a Galois field $\mathbb{F}_q$. For simplicity, it is considered that each packet is a single symbol in $\mathbb{F}_q$.

\begin{dfn}
A packet \pma{n} corresponds to $n$'th packet that has entered the transmission queue.
 A packet \pma{n} is older than \pma{m} if $n<m$, otherwise it is newer. 
\end{dfn}

\begin{dfn}\label{def:decod pk}
A packet \pma{n} is decoded by a user \um{i} if the individual value of \pma{n} has been revealed by applying Gaussian elimination on the already received network coded packets. 
\end{dfn}

\begin{dfn}\label{def:delivery p}
	A packet \pma{n} is \textit{delivered} to a user if all older packets	$\pp{1},\pp{2},\ldots,\pp{n-1}$ have been decoded by that user. The number of delivered packets by \um{i} at time slot $t$ is shown with \dm{i}.
\end{dfn}

\begin{dfn}\label{def:seen pk}
	The user \um{i}   has \textit{seen} a packet \pma{n},  if it can compute a linear combination of the form 
	$(\pp{n}+\boldsymbol{q})$, where $\boldsymbol{q}$ is a linear combination of the packets older than $\pp{n}$.
\end{dfn}

\begin{exm}
Consider Table \ref{buffer example} as an example of a user's buffer. By Definitions  \ref{def:decod pk}, \ref{def:delivery p} and \ref{def:seen pk} $\pp{1},\pp{2},\pp{3}$ are delivered packets thus they are decoded and seen too. $\pp{5}$ is decoded and seen however, it is not delivered. $\pp{6}$ is just a seen packet and $\pp{4}$ is neither a seen packet  nor decoded.	
\end{exm}

\begin{table}[bh]
	\caption{An example of a user's buffer} \label{buffer example}
	\centering 
	\begin{tabular}{|c|c|c|c|c|c|c|c|}
		\hline 
		$\pp{1}$ & $\pp{2}$ & $\pp{3}$ & \ldots &$\pp{5}$ & $\pp{6}+\pp{4}$ & \ldots \\ 
		\hline 
	\end{tabular} 
\end{table}

  Seeing, decoding and delivering are the situations of a packet in the users' buffers with different level of strength. Note that a delivered packet is also a decoded and seen packet and a decoded packet is a seen packet but the opposite is not true necessarily. Seeing a packet is an important concept for the queue management and decoding process  \cite{2009sundararajanfeedback}. When all the users have seen a packet, that packet is dropped from the transmission queue and the users save it until they receive older packets to decode it.

\begin{dfn}
At time $t$, the next required packet of \um{i} is the oldest unseen packet in its buffer, and it is denoted by \nm{i}.
\end{dfn}

 \begin{exm}
In Table \ref{buffer example} the next required packet will be $\pp{4}$.
 \end{exm}

 There is full feedback from the users to the transmitter so that in each time slot the sender has complete information about what packets the users have correctly received or lost and their next required packets. The sender uses this information to determine the combination of the packets for the next transmission \cite{2009sundararajanfeedback}.
 
  \begin{dfn}
  A transmission  $s(t)$ is a symbol in $\mathbb{F}_q$ and comprises the next required packets of the users along with the coding coefficients at time slot $t$, which is given by
  	\begin{equation}
 	s(t)=\sum_{i=1}^{\nu}{\alpha_i(t) \nn{i}},
 	\end{equation}
 	where $\alpha_i(t)$s are chosen from $\mathbb{F}_q$ using a non-block-based coding scheme, which will be defined shortly.
 \end{dfn}
 
 \begin{dfn}
 	A transmission $s(t)$ is innovative for \um{i}, if it cannot be computed from the information stored in the buffer of \um{i}.
 \end{dfn}
 
\begin{dfn}
The delivery rate  of a user \um{i} is given by $\rate{i}(t)=\frac{\dd{i}}{t}$, and the average rate at which packets are delivered to the user is  $\rate{i}=\displaystyle\lim_{t\rightarrow\infty}{\rate{i}(t)}$.
\end{dfn}

\subsection{Coding scheme}
The sender employs linear network coding that was proposed in \cite{2009sundararajanfeedback}. The method of packet encoding is given in Algorithm \ref{alg:codscheme} \cite{2009sundararajanfeedback}. In each time slot, the transmitter makes a list of the next requested packets \pma{j} by the users in descending order of the packet indices, excluding those users whose required packets have not yet arrived  into the transmission queue. Let $\mathcal{G}_j$ be the group of the users whose next requested packet is \pma{j}. Starting with group $\mathcal{G}_j$ with  the highest index, it will add the packet \pma{j} into $s(t)$ only if the user(s) in $\mathcal{G}_j$  do not otherwise receive an innovative packet. Furthermore, to ensure that  we can always find an innovative transmission for all the users using this coding scheme, the field size should be $q\geq \nu$ \cite{2009sundararajanfeedback}. To check if $s(t)$ is innovative, \textit{Gaussian elimination} of $s(t)$ and the information of the buffers of \um{i}s is used. For each $\uu{i}\in \mathcal{G}_j$ the residual of Gaussian elimination $r_i$ is stored in a set which is called the \textit{veto list}. By subtracting the veto list from the field, it is ensured that the  chosen coefficient for \pma{j} makes it possible for all
users in $\mathcal{G}_j$ to decode $\pp{j}$. 

 \begin{algorithm}[t]
 	\begin{algorithmic}[1]
 		\STATE Organize users $\uu{1},\cdots,\uu{\nu}$ into groups $\mathcal{G}_j$, so that $\mathcal{G}_j$ contains  at least one user.  
 		\STATE Initialize $s(t)=0$.
 		\FOR {each group $\mathcal{G}_j $, from high	to low $j$,}
 		\STATE Initialize the empty \emph{veto list} $v_j=\left\{\right\}$.
 		\FOR {each user $\uu{i}\in \mathcal{G}_j$}
 		\STATE Calculate $\rho_i$, the residual of performing Gaussian elimination on $s(t)$ with the {transmissions stored} in $\uu{i}$'s buffer.
 		\IF {$\rho_i=0$}
 		\STATE $v_j \leftarrow v_j \cup \{0\}$.
 		\ELSIF {$r_i=\alpha \pp{j}$ for some field element $\alpha$}
 		\STATE $v_j \leftarrow v_j \cup \{\alpha\}$.
 		\ENDIF
 		\ENDFOR
 		\IF {$0 \in v_j$}
 		\STATE {$a_j \triangleq \min(\mathbb{F}_q \backslash v_j)$.} %to the smallest value from $\mathbb{F}_M$ not listed in $v_j$.
 		\STATE Set $s(t)=s(t)+a_j\pp{j}$.
 		\ENDIF
 		\ENDFOR
 	\end{algorithmic}
 	\caption{Coding algorithm \cite{2009sundararajanfeedback}.} \label{alg:codscheme}
 \end{algorithm}

\begin{table*}[h]
		\caption{An example of the coding scheme and transmission process}\label{tab:coding}
	\begin{center}	
		\begin{tabular}{|c|c|c c|c c|c c|}
			\hline 
			$t$&$0$&\multicolumn{2}{|c|}{$1$}&\multicolumn{2}{|c|}{$2$ }&\multicolumn{2}{|c|}{$3$}\\ 
			\hline 
			$\text{Source buffer}$&
			$\pp{1}$-$\pp{10}$&\tick&$\pp{1}$-$\pp{11}$&\cross&$\pp{1}$-$\pp{11}$&\cross&$\pp{1}$-$\pp{11}$\\ 
			\hline 
			$s(t)$&  &&$\pp{11}$&& $\pp{11}+\pp{6}$ &&$\pp{6}$\\ 
			\hline 
			$U_1$ \text{buffer}&$\pp{1}$-$\pp{10}$&\cross&$\pp{1}$-$\pp{10}$&\tick&$\pp{1}$-$\pp{11}$&\tick& $\pp{1}$-$\pp{11}$ \\
			\hline 
			$U_2$ \text{buffer}&$\pp{1}$-$\pp{5}$,$\pp{9}$&\tick& $\pp{1}$-$\pp{5},\pp{9},\pp{11}$ &\cross&$\pp{1}$-$\pp{5},\pp{9},\pp{11}$&\tick&$\pp{1}$-$\pp{6},\pp{9},\pp{11}$\\ 
			\hline 
			$U_3$ \text{buffer}&$\pp{1}$-$\pp{2}$,$\pp{6}+\pp{3}$&\cross& $\pp{1}$-$\pp{2}$,$\pp{6}+\pp{3}$&\cross&$\pp{1}$-$\pp{2}$,$\pp{6}+\pp{3}$&\tick&$\pp{1}$-$\pp{3},\pp{6}$\\ 
			\hline 
		\end{tabular} 
	\end{center} 
	\vspace{2mm}
\caption*{Ticks and crosses in the Source buffer row represent that a new packet has been entered the transmission queue or not respectively, while the users' rows represent their channel states (successful and unsuccessful reception, respectively) at each transmission.}
\end{table*}

%	\vspace{-2mm}
\begin{exm}
\label{ex-coding}
Consider a system with three users such that $\cc{1}>\cc{2}>\cc{3}$.  For simplicity, let $N_1(0)>N_2(0)>N_3(0)$ so that there is only a single user in each group $\mathcal{G}_j$ corresponding to the next required packet $\pp{j}$. An example of the transmission scheme is given in Table \ref{tab:coding}.

 At $t=0$, $10$ packets have arrived into the transmission queue. $U_1$ has $\pp{1}$-$\pp{10}$ as delivered packets and $\pp{1}$-$\pp{5}$ have been delivered to $U_2$ and also it has received $\pp{9}$. $U_3$ has $\pp{1},\pp{2}$ as delivered packets and it has received the combination $\pp{6}+\pp{3}$.
  We have $d_1(0)=10$, $d_2(0)=5$ and $d_3(0)=2$. The next required packets of the users are $N_1(0)=\pp{11}$, $N_2(0)=\pp{6}$ and $N_3(0)=\pp{3}$. At $t=1$, the sender checks the next required packets of the users and starts encoding by the highest index packet $\pp{11}$. The sender sets $s(1)=\pp{11}$ at the first step, then because $s(1)$ is innovative for all the users, it is sent without adding any other packet to it. After the  transmission of $s(1)$, $U_2$ receives the packet successfully while $U_1$ and $U_3$ have erasures. At the next time slot $t=2$, again $\pp{11}$ is the next required packet with the highest index and encoding starts with $s(2)=\pp{11}$. After that, the sender checks the users' buffer information and finds out that $\pp{11}$ is not innovative for $U_2$, thus it adds $N_2(2)=\pp{6}$ to $s(2)$. Now $s(2)$ is innovative for all the users and it can be transmitted. After transmission $U_1$ receives the packet and decodes $\pp{11}$, but $U_2$ and $U_3$ cannot receive the packet.
At $t=3$, the required packet with the highest index is $N_1(3)=\pp{12}$. However it has not entered the transmission queue, so $U_1$ is not considered for encoding. Therefore, the encoding starts with $N_2(3)=\pp{6}$ and it is innovative for $U_2$ and $U_3$ thus it is transmitted. All the users receive this packet. It is not innovative for $U_1$ but $U_2$ receives its needed packet and $U_3$ uses it to decode the combination $\pp{6}+\pp{3}$ and reveal $\pp{3}$ as its required packet.
\end{exm}

\begin{dfn}
 A user with the highest next requested packet index \nm{i} in $s(t)$ is named the \textit{leader}, also this transmission is called a \textit{leader transmission} for \nm{i}. In Example \ref{ex-coding}, \um{1} is  the leader at time slots $t=1,2$ and $U_2$ is the leader at $t=3$. Thus, the transmissions  are leader transmissions for $\pp{11}$ in $t=1,2$ and for $\pp{6}$ in $t=3$.
\end{dfn}

\begin{dfn} \label{def: diffknowledge two}
At time slot $t$,  we call $s(t)$  a  \textit{differential knowledge} \footnote{In \cite{futhesis} \textquotedblleft{knowledge differential}\textquotedblright was used for this concept and the definition was somewhat different. Here, we have streamlined the definition as applicable to the model of this paper and \cite{2009sundararajanfeedback}.} transmission for a user \um{j} if $s(t)$ leads to the delivery of its next required packet, otherwise it is called a non-differential transmission. If $U_i$ is the leader and $s(t)$ is a differential knowledge for $U_j$, we say that $U_j$ has a differential knowledge from $U_i$ and the probability of this event is shown by $D_j^i$. In each time slot, the probability of transmitting differential knowledge for the leader, $D_i^i$, is one (In fact, leader transmission is a differential knowledge for the leader). 
\end{dfn}

\begin{exm}
	In  Table \ref{tab:coding}, $s(1)$ and $s(2)$  are leader and differential knowledge   transmissions for \um{1}, while $s(2)$ is a differential knowledge transmission for \um{2}. Moreover, $s(3)$ is a leader transmission (and also differential knowledge) for $\uu{2}$. Thus, if they receive these corresponding transmissions their next required packets are delivered. Note that $s(1)$ is a non-differential transmission for $\uu{2}$ and all transmissions are non-differential for $\uu{3}$.
\end{exm}

%% file: 3_Delivery_Rate_Analysis.tex
\section{Delivery Rate Analysis}
\label{sec:Delivery Rate Analysis}
In this section, the analysis of the delivery rate is proposed. First, let us introduce the following sets:
\begin{itemize}
	\item $\mathcal{S}(t)$: Set of  packets arrived at the sender untill end of time slot $t$. The size of $\mathcal{S}(t)$ is denoted by  $|\mathcal{S}(t)|$.
	\item $\mathcal{S}_i(t)$: Set of  packets that  user $\uu{i}$ has seen untill end of time slot $t$. 
\end{itemize}

Based on the coding scheme \ref{alg:codscheme}, the sender drops a packet from the transmission queue when all the users have seen that packet. Thus, the size of the transmission queue is given by
\begin{equation}
\mathcal{Q}(t)=|\mathcal{S}(t)|-|\displaystyle\cap_{i=1}^{\nu}\mathcal{S}_i(t)|.
\end{equation}

Similar to the physical transmission queue, we can define a virtual queue for each $U_i$ as the difference of $\mathcal{S}(t)$ and $\mathcal{S}_i(t)$.  The size of such a virtual queue is given by
\begin{equation}
\mathcal{Q}_i(t)=|\mathcal{S}(t)|-|\mathcal{S}_i(t)|.
\end{equation}

$\mathcal{Q}_i(t)$ can be modeled by a Markov chain  (Fig. \ref{Asymmetric Markov}). The transition between the states depends on both the incoming packets to the virtual queue and  seeing the packets by the user.
If $c_i>\lambda$ the Markov chain is positive recurrent and has steady states and if $c_i \leq \lambda$ it is transient.

We can use the steady state distribution for the positive recurrent case, however, for the users with a transient Markov chain we use another approach. Thus, we separate users in two groups i.e., $\hh$ and $\LL$ due to the different behavior of their Markov chains. Accordingly, we use the following notation

\begin{itemize}
\item $\mathcal{H}=\{\uu{i}   : c_i > \lambda \} $,
\item $\mathcal{L}=\{\uu{i}   : c_i \le \lambda \} $,
\item $|\mathcal{H}|=\nu_h$.
\end{itemize}

\begin{figure}[]
	\centering
	\begin{tikzpicture}[]
	%Define standard arrow tip
	\tikzset{>=stealth'	}
	\tikzstyle{ann} = [draw=none,fill=none,right]
	\tikzstyle{circ}=[fill=blue!20,minimum width=1cm,draw,circle]
	
	\node (S0) [circ,align=left] at (0,0){$0$};
	\node (S1) [circ,right of=S0,node distance=2.2cm,align=left]  {$1$};
	\node (S2) [circ,right of=S1,node distance=2.2cm,align=left]  {$2$};
	\node (S3) [ann,right of=S2,node distance=1.2cm,align=left]  {\huge{$\dots$}};
	
	\path 
	(S0) edge [->,loop left,every loop/.style={min distance=1cm,in=-155,out=155,looseness=10}] node {$\bar\lambda + \lambda c_i$} (S0)
	edge [->,bend left] node[above] {$\lambda\bar{c_i}$} (S1)
	(S1) edge [->,loop below,every loop/.style={min distance=1cm,in=-115,out=-65, looseness=10}] node {$\bar\lambda\bar c_i + \lambda c_i$}(S1)
	edge [->,bend left] node[below] {$\bar\lambda c_i$}(S0)
	(S2) edge [->,loop below,every loop/.style={min distance=1cm,in=-115,out=-65 ,looseness=10}] node {$\bar\lambda\bar c_i + \lambda c_i$} (S2)
	edge [->,bend left] node[below] {$\bar\lambda c_i$} (S1)
	(S1) edge [->,bend left] node[above] {$\lambda\bar{c}_i$} (S2);
	\end{tikzpicture}
	\caption{Markov chain for the size of the virtual queue $\mathcal{Q}_i(t)$.}
	\label{Asymmetric Markov}
\end{figure}

For $U_i \in \mathcal{H}$ , the steady state distribution of $\mathcal{Q}_i(t)$ with the Markov chain in Fig. \ref{Asymmetric Markov} is given by \cite{2014Sadeghidynamic}
\begin{equation}
\label{eq:state}
\pi_i(n)= \left(1-\frac{\lambda\bar{c}_i}{\bar{\lambda}c_i} \right) \left(\frac{\lambda\bar{c}_i}{\bar{\lambda}c_i} \right)^n.
\end{equation}

	For $1\leq i\leq \nu_h$, when $\mathcal{Q}_i(t)=0$, $U_i \in\hh$ is in the zero state and all the packets in the sender have been seen by $U_i$. Thus, all of them are decoded and delivered. In this situation, if no new packet arrives at the sender, the transmitted packet has no new information for $U_i$. The probability of this event is $\bar{\lambda}\pi_i(0)$,  otherwise, by the innovation guarantee property, the transmitted packet has new information for the user  (i.e., if the user receives  this transmission, it can see a new packet) with probability
		\begin{equation}
	\label{eq:zerostate}
	1-\bar{\lambda}\pi_i(0)=\frac{\lambda}{c_i}.
	\end{equation}

\begin{theorem}
	\label{theorem:Hdelivery}
	In steady state, the delivery rate of each $\uu{i}\in\mathcal{H}$ asymptotically tends to $\lambda$.
\end{theorem}

\begin{proof}
		As $t \rightarrow\infty$, the size of $\mathcal{S}(t)$ tends to $\lambda t$.  Using (\ref{eq:zerostate}), the transmitted packets have new information for a $\uu{i}\in\mathcal{H}$ in $(\frac{\lambda}{c_i})t$ fraction of time and  $\uu{i}$ receives  $\frac{\lambda}{c_i}tc_i=\lambda t$ number of these transmissions on average. Because all of these received  transmissions are innovative, $\uu{i}$  have seen all the packets and it can decode them, so the packets are delivered. Thus, the asymptotic delivery rate of the users in $\hh$ will be 
	\begin{equation}
	\lim_{t\rightarrow\infty}{\frac{d_i(t)}{t}}=\frac{\lambda t}{t}=\lambda.
	\end{equation}
	\end{proof}
The asymptotic delivery rate of the users in $\hh$ is given by Theorem \ref{theorem:Hdelivery}. To estimate the delivery rate of the users in $\LL$, we need the following arguments on the leadership probability and differential knowledge.

\subsection{Probability of becoming the leader}\label{Probability of becoming the leader} 
In each time slot, at least a user is the leader (i.e., the next required packet is the highest index packet in $s(t)$), and encoding is done based on its requested packet.  In the following, we analyze the probability of  the leadership for the users in $\hh$ and $\LL$.

\subsubsection{$\hh=\varnothing$}\label{H is empty}
If $\hh=\varnothing$, all the users are in $\LL$ with the link capacities smaller than $\lambda$. After some time slots, users are left behind the transmission queue and their next required packets always exist at the sender. On the other hand, the first user with the highest link capacity $U_1\equiv U_l$  receives more packets, so that after some time slots its next requested packet has  the highest index in $s(t)$. 
In this case, by ignoring some time slots from the beginning of the transmission, $ U_1$ is always the leader and its leadership probability is assumed to be 1.

\subsubsection{$|\hh|=1$}\label{H=1}
In this case, the strongest user $\uu{1}$ is the only member of $\hh$. 
Because $c_1>\lambda$, there are some time slots that $N_{1}(t)$ has not arrived at the sender. Using (\ref{eq:zerostate}), the probability of this event is $1-\frac{\lambda}{c_1}$. In these time slots, because the sender has no new information for $\uu{1}$, the second user $\uu{2}\equiv U_l$ is the leader. Note that the only user in $\LL$ that can be the leader is $U_l$ based on the argument in the previous  item. Thus, in this case, there are only two leaders $U_1$ and $U_2$.  $U_1$ is the leader with probability of $\frac{\lambda}{c_1}$ and $U_2$ is the leader with the complement probability.

\subsubsection{$|\hh|>1$}\label{H>1}
In this case, since all the users in $\hh$ have capacities greater than $\lambda$, there are situations where more than one user are leaders at the same time. Thus, analysis of the leadership probability is not trivial. However, we can make an estimation of the $U_l$ leadership probability and the probability that the leader is in $\hh$.

Based on the Markov chain analysis, each user in $\hh$ comes back to the zero state in steady state. It can happen for more than one user or even all the users in $\hh$ at the same time. When all the users in $\hh$ are in the zero state and no new packet arrives at the sender, $U_l$ will be the leader based on the previous arguments. Now we estimate the probability of $U_l$ leadership in this case.

Let us define a new virtual queue as 
\begin{equation}
\mathcal{Q}_h(t)=|\mathcal{S}(t)|-|\displaystyle\cap_{i=1}^{\nu_h}\mathcal{S}_i(t)|.
\end{equation}
When all the users in $\hh$ are in the zero state $\mathcal{Q}_h(t)=0$.
Since each $\mathcal{S}_i(t)$ is a subset of $\mathcal{S}(t)$, we have \cite{2008sundararajanarq} 
\begin{equation}
\label{eq:setineq}
|\mathcal{S}(t)|-|\cap_{i=1}^{\nu_h}\mathcal{S}_i(t)| \leq \sum_{i=1}^{\nu_h}({|\mathcal{S}(t)|-|\mathcal{S}_i(t)|}),
\end{equation}
thus,
\begin{equation}
\label{eq:queuineq}
\mathcal{Q}_h(t) \leq \sum_{i=1}^{\nu_h} {\mathcal{Q}_i(t)}.
\end{equation}
%%%%%%%%%%%%%%%%%%%%%%%%%%%%%%%%%%
Using Markov chain, the steady state expected size of the virtual queue is given by
\begin{equation}
\label{eq:expectation}
\displaystyle\lim_{t \rightarrow \infty} \mathbb{E}[\mathcal{Q}_i(t)]= \sum_{n=0}^{\infty}{n\pi_i(n)}=\frac{\bar{c}_i\lambda}{c_i-\lambda}.
\end{equation}
We consider a Markov chain for $\mathcal{Q}_h(t)$ and a corresponding parameter $c_h$ as $c_i$ in $\mathcal{Q}_i(t)$. With this consideration and taking the expectation on both sides of (\ref{eq:queuineq}), we obtain
\begin{equation}
\label{eq:chexpect}
\frac{\bar{c}_h\lambda}{c_h-\lambda}    \leq	\sum_{i=1}^{\nu_h}{\frac{\bar{c}_i\lambda}{c_i-\lambda}}.
\end{equation}
Using (\ref{eq:chexpect})  we have
\begin{equation}
\label{eq:chmin}
\frac{1+\lambda \eta}{1+\eta} \leq c_h,
\end{equation}
where, $\eta=\sum_{i=1}^{\nu_h} {\frac{\bar{c_i}}{c_i-\lambda}}.$

The probability of $\uu{l}$ leadership is $\bar{\lambda}\Pr(\mathcal{Q}_h=0)$. 
Using (\ref{eq:chmin}) and (\ref{eq:zerostate}), if we set
\begin{equation}
\label{eq:ch}
\frac{1+\lambda \eta}{1+\eta}=c_h,
\end{equation}
 an upper bound for $\bar{\lambda}\Pr(\mathcal{Q}_h=0)$ will be  given by
\begin{equation}
\label{eq:chtime}
\bar{\beta}_h=1-\frac{\lambda}{c_h}.
\end{equation}
$\bar{\beta}_h$ is a possible minimum fraction of the time that $U_l$ is the leader. Consequently, the fraction of the time that the leader is in group $\hh$ is given by the complement probability $\beta_h$.

\subsubsection{Leadership probability model}
First, we define the following parameter to summarize the  previous  results.
\begin{equation}
\label{eq:beta}
\beta\triangleq 
\begin{cases}
\beta_h & \mathcal{H}\neq\varnothing, \\
0 & \mathcal{H}=\varnothing.
\end{cases}
\end{equation}
It shows the leadership probability of the users in $\hh$. If $\hh\neq\varnothing$, using (\ref{eq:chtime}) the leader is in group $\hh$ with the probability of $\beta_h$, so $\beta=\beta_h$. Note that if $|\hh|=1$ we have $c_h=c_1$. For the case of $\hh=\varnothing$, there is no user in $\hh$, so $\beta=0$ and $U_l$ is the leader with probability of $\bar{\beta}=1$.

In the  case where $|\hh|>1$, due to the simultaneity of the leadership for the users in $\hh$ in some time slots, analysis of leadership probability for each individual user in this group is very complicated. However, to estimate the delivery rate of the users in $\LL$,  we need the leadership probability. Thus, we omit exact analysis and use (\ref{eq:ch}) and (\ref{eq:chtime}) to make an estimation for differential knowledge and delivery rate of the users in $\LL$.

Suppose a virtual user $U_h$ with capacity $c_h$ corresponds to $\mathcal{Q}_h$. To estimate differential knowledge and delivery rate of the users in $\LL$, we consider $U_h$ as the representative of the users in $\hh$. Using (\ref{eq:beta}) the leader is in $\hh$ in $\beta$ fraction of the time and is $U_l$ with the complement probability $\bar{\beta}$. When one or some of the users in $\hh$ are the leaders, we consider $U_h$ as the leader and we use $c_h$ in our analysis instead of the capacity of the users in $\hh$. Although it is not an exact model, it  helps us to make a tractable model for the delivery rate of the users in $\LL$. We summarize the above argument with the following approximation.

\begin{app}\label{app-RH}
In the rest of this paper, we consider $U_h$ as the representation for the leadership of the users in $\hh$. We say $U_h$ is the leader if any user in $\hh$ is the leader. We also use (\ref{eq:ch}) in our analysis instead of the capacity of the users in $\hh$.
Thus, for simplicity we consider only two leaders $U_h$ and $U_l$. $U_h$ is the leader with probability of $\beta$ and $U_l$ with complement probability $\bar{\beta}$.
\end{app}
In the following, the indices $h$ and $l$ will be used for the corresponding parameters for $U_h$ and $U_l$, respectively. For instance, $d_h(t)$ and $d_l(t)$ are used for the number of delivered packets to $U_h$ and $U_l$ at time slot $t$, respectively

\subsection{Probability of differential knowledge for the users in $\LL$} \label{sec: Probability of differential knowled}

In each time slot, there is just a leader $U_h$ or $U_l$ and by Definition \ref{def: diffknowledge two}  at least one user can receive a differential knowledge from one of these leaders.
A differential knowledge transmission for \um{i} is sent in a time slot $t$, if $N_i(t)$ has been encoded in the transmission packet. When $U_h$ is the leader the encoding is started with its requested packet  and if it is the first request of a packet $\pp{n}$ by $U_h$ it will be the first transmission of $\pp{n}$ and we have $s(t)=\pp{n}$. The transmission of $\pp{n}$ will continue until it is received by $U_h$ and during this process, the requested packet of another user $U_i$ is added to $s(t)$ if it receives $\pp{n}$ while $U_h$ has not received it. Here, $U_i$ has a differential knowledge from $U_h$.
 When $U_l$ is the leader there is a similar explanation, however, note that only the users in $\LL$ can have differential knowledge from $U_l$. 

 Before calculating the probability of differential knowledge, the probability of the leader transmissions for \um{h} and \um{l} are determined.
$L_h^*(k)$ and $L_l^*(k)$  are the probabilities of $k$ unsuccessful leader transmissions of a requested packet by $U_h$ and $U_l$, respectively. 
$L_h(k)$  is  the probability of $k$ leader transmissions of a delivered packet by $\uu{h}$. Remember that we replaced $\uu{h}$ for the all members in group $\hh$.

We start by deriving $L_h^*(k)$. When the transmission of a new packet for $U_h$ is started, it will continue until the packet is received. Note that transmissions for the other users have no new information for $U_h$ and they have no effect on  $L_h^*(k)$, so it is given by
\begin{equation}
L_h^*(k)={\bar{c}_h}^kc_h,
\label{eq:LH*}
\end{equation}
where $\bar{c}_h^k$ is probability of $k$ unsuccessful leader transmissions of a packet for $U_h$.

To determine $L_l^*(k)$, note that $U_l$ can deliver a packet by the leader transmissions or  differential knowledges from $U_h$. The probability of  the leader transmission for $U_l$ is $\bar{\beta}_h$ and the probability of receiving differential knowledge is $\beta D_l^hc_l$.  We normalize the probabilities to $\bar{\beta}_h+\beta D_l^hc_l$, in order to restrict our probability space to these events.
Accordingly, $L_l^*(k)$ is given by
\begin{equation}
L_l^*(k)=\left(\frac{\bar{\beta}\bar{c}_l}{\bar{\beta}+\beta D_l^hc_l}\right)^k\left(\frac{\bar{\beta}{c}_l+\beta D_l^hc_l}{\bar{\beta}+\beta D_l^hc_l}\right).
\label{eq:LL*}
\end{equation}

To calculate $L_h(k)$, we consider two cases of $k=0$ and $k>0$. With the given Definition of $L_h(k)$, $k=0$ denotes that $U_h$ received a packet using no leader transmission.  This would mean  $U_h$ received a differential knowledge transmission which is not possible by our model, so $L_h(0)=0$. For $k>0$, to have exactly $k$ leader transmissions there should be $k-1$ unsuccessful leader transmissions followed by a successful one. Thus we have
\begin{equation}
L_h(k)=\begin{cases}
0,&k=0,\\
\bar{c}_h^{(k-1)}c_h,&k\neq{0}.
\end{cases}
\label{eq:LH}
\end{equation}

Now the probability of differential knowledge for a user $U_i$ is calculated using the complement of the probability that it has not seen the required packet of the leaders, i.e.,
\begin{align}
\label{eq:DH}
&D_i^h=
1-\left(\sum_{k=0}^{\infty}{L_h^*(k)\bar{c}_i^k}\right),
\end{align}
\begin{align}
\label{eq:DL}
D_i^l=\begin{cases}
1,&\uu{i}=\uu{l},\\
1-\left(\displaystyle{\sum_{k=0}^{\infty}{L_l^*(k)\bar{c}_i^k}}\right),&\hh=\O{},\\
1-\left(\displaystyle{\sum_{k=0}^{\infty}{L_l^*(k)\bar{c}_i^k}}\right) \times \\  \hspace{2.5mm}\quad\left(\displaystyle{\sum_{k=0}^{\infty}{L_h(k)\bar{c}_i^k}}\right),&\hh \neq \O{}.
\end{cases}
\end{align}
Note that the summation in (\ref{eq:DH}) represents the probability that $\uu{i}$ has not seen the requested packet by \um{h} while \um{h} has been the leader.
Similarly, in (\ref{eq:DL}) when $\hh =\O{}$, the summation is the same probability while \um{l} has been the leader.  In the case of $\hh \neq \O{}$,  there is the second summation which is the probability of transmitting the packet prior to the leadership of \um{l} and they are independently multiplied.
 
\subsection{Delivery rate} \label{Delivery rate}
\begin{theorem}\label{theorem1}
	The asymptotic delivery rate of the users in this system is given by
	\begin{equation}
	R_{i}=\begin{cases}
	\lambda, & U_i\in{\hh},\\
	\displaystyle{\frac{\left({\beta}D_i^h+{\bar{\beta}}D_i^l\right)\cc{i}}{1-B_i}},
	& U_i\in{\LL}.
	\end{cases}
	\label{eq:Ri}
	\end{equation}
In the above, we have
	\begin{equation}
	\label{eq: B_i}
	B_i=\frac{{\beta}\bar{D}_i^hc_i}{\lambda}+ \frac{ {\bar{\beta}} {\bar{D}_i^l}c_i}{R_{l}}.   	
	\end{equation}
	\end{theorem}

\begin{cor}
	The coding method in Algorithm \ref{alg:codscheme} can achieve multirate packet delivery in the described system model to the users in $\LL$ whenever it is non-empty.
\end{cor}

According to (\ref{eq:Ri}), in a system with non-empty set $\LL$, users experience different delivery rates. On the other hand, although users in $\hh$ have equal delivery rate, users in group $\LL$ experience delivery rates proportional to their capacity. This  is the desired multirate packet delivery property of the coding method.

\begin{proof}
The average delivery rate of  \um{i} is given by $R_i=\displaystyle\lim_{t\rightarrow\infty}\dd{i}/t$. 
If $U_i \in \mathcal{H}$ the asymptotic delivery rate  is $\lambda$ using Theorem \ref{theorem:Hdelivery}.  For the users in $\LL$, the value of \dm{i} is estimated by accumulation of  the number of the packets  delivered to \um{i} from different types of transmissions. \um{i} receives differential knowledge  from users in $\hh$  with the probability ${\beta}D_i^hc_i$, and from $\uu{l}$ with the probability $\bar{\beta}D_i^lc_i$. Furthermore, there are packets that \um{i} receives from non-differential transmissions, which are distributed in the buffer of \um{i} between \pma{1} to $\pp{d_h(t)}$  and \pma{1} to $\pp{d_l(t)}$. 
Assuming that these packets are uniformly distributed, we have
%\begin{align}
%\label{eq:d_i(t)}
%\dd{i}=&{\beta}D_i^hc_it+\bar{\beta}D_i^lc_it+
%{\beta}\bar{D}_i^hc_i\frac{\dd{i}}{\dd{h}}t+\\ \nonumber &{\bar{\beta}} {\bar{D}_i^l}c_i\frac{\dd{i}}{\dd{l}}t,
%\end{align}
\begin{equation}
\label{eq:d_i(t)}
\dd{i}={\beta}(D_i^hc_i+\bar{D}_i^hc_i\frac{\dd{i}}{\dd{h}})t+\bar{\beta}(D_i^lc_i+{\bar{D}_i^l}c_i\frac{\dd{i}}{\dd{l}})t,
\end{equation}
 where ${\beta}\bar{D}_i^hc_i\frac{\dd{i}}{\dd{h}}t$ is the fraction of the received packets from non-differential transmissions while \um{h} is the leader. Similarly ${\bar{\beta}} {\bar{D}_i^l}c_i\frac{\dd{i}}{\dd{l}}$ is the received non-differential packets while \um{l} is the leader. These packets are in the delivered region of \um{i}'s buffer $\pp{1}$ to $\pp{d_i(t)}$. From (\ref{eq:d_i(t)}) we have
 \begin{equation}
 \label{eq:del before}
 	 \dd{i}={\beta}D_i^hc_it+\bar{\beta}D_i^lc_it+
 	d_i(t)B_i.
 \end{equation}
 Using (\ref{eq:d_i(t)}) and (\ref{eq:del before}) $B_i$ is given by (\ref{eq: B_i}). Note that $\displaystyle\lim_ {t\rightarrow\infty}\frac{d_h(t)}{t}=\lambda$.
 Finally, using  (\ref{eq:del before}) the delivery rate is given by (\ref{eq:Ri}).
 \end{proof}
 

%% file: 4_Delay_Analysis.tex
\section{Delay Analysis} \label{sec:Delay Analysis}
Different types of delay analysis have been studied in \cite{2009sundararajanfeedback,2012SadeghiDeliverydelayanalysis,2014Sadeghidynamic} for this 
coding scheme in homogeneous networks. They considered both the decoding delay and delivery delay. However, in such a system where the packets can be used only if they are delivered, the decoding delay is less important than the delivery delay; because it is possible that a user decodes a packet but it must wait until it is actually delivered in order to the application layer.  In the literature, the delivery delay has been considered as the time between when a packet enters  the transmission queue and its delivery to the application at each user \cite{2009sundararajanfeedback}. Using this definition of the delivery delay in heterogeneous networks, there may be a large difference between  delivery times of a packet for different users. On the other hand, weak users left behind from the transmission queue, still seek older packets to complete their delivery. Therefore, we study the delivery delay using a new definition, which is based on the time that each user waits for a packet after it is first requested by that user. Using this new definition, the delivery delay  of the users is measured independently of each other and with respect to their capability of delivering packets. We believe that this new definition is more suitable for heterogeneous networks and moreover, it leads to a closed form for the delivery delay.

\begin{dfn}
The delivery delay of a packet for a user \um{i} is the time between the first request of that packet and its delivery  which is shown by $\theta_i$.
\end{dfn}

\begin{theorem}
Suppose that  $d_{u_i}$  is the probability of delivering a packet by $U_i$ in each time slot.  For $U_i$, the probability that a packet has   $T$ time slots delivery delay $\theta_i=T$ is given by
\begin{equation}
\Pr(\theta_i=T)=\begin{cases}
{d_{u_i}^2(1-d_{u_i})^{(T-1)}}/{R_i},&T>0,\\
1-\displaystyle\sum_{T=1}^{\infty}{\Pr(\theta_i=T)},&T=0.
\end{cases}
\label{eq:delivery delay}
\end{equation}

\end{theorem}

\begin{proof}
A packet is delivered to a user $U_i$ with the probability of $d_{u_i}$, then $U_i$ requests the next packet and it can deliver that packet after $T>0$ time slots with the probability of  $(1-d_{u_i})^{(T-1)}d_{u_i}$.
Therefor, the number of packets with  delivery delay of $T>0$ is $t{d_{u_i}^2(1-d_{u_i})^{(T-1)}}$, and $\Pr(\theta_i=T)$ is given by
$\displaystyle\lim_{t\rightarrow\infty}{td_{u_i}^2(1-d_{u_i})^{(T-1)}}/{d_i(t)}$. By summation on $\Pr(\theta_i=T)$ for $T>0$, all packets with the non-zero delivery delay in \um{i} buffer are considered
and the probability of the rest of them is given by complement probability that is given in (\ref{eq:delivery delay}). These are the packets with the zero delivery delay which have been decoded sooner than the previous packets. When the user  needs them they have already been delivered and hence are not requested again from the sender.
\end{proof}

To calculate $\Pr(\theta_i=T)$, the value of $d_{u_i}$ is needed that could be complicated in some cases. For instance, if $|\hh|>1$, calculation of $d_{u_i}$ is rather complicated for the users in $\hh$ . 
In the following, we  compute $d_{u_i}$ for some special cases. 

\subsection{ $\hh=\varnothing$}\label{No user in group H}
In this case, the strongest user is always the leader and the other ones receive their packets only via differential knowledge transmissions. Thus, according to Section \ref{Delivery rate} the probability of packet delivery is given by
\begin{equation}
d_{u_i}=\begin{cases}
c_1,&i=1,\\
D_i^1c_i,&i>1.
\end{cases}
\label{NO_R_delay}
\end{equation}

For $U_1$ (the strongest user), $d_{u_1}$ is the same as the channel capacity because it is the strongest user and all the packets in its buffer are assumed to be delivered. However, for the other users $d_{u_i}$ is different, since they also receive  non-differential packets which affects the number of delivered packets and the delivery rate, while $d_{u_i}$ is the probability of receiving a requested packet and its delivery at the same time.
Now, $\Pr(\theta_i=T)$ can be determined using (\ref{eq:delivery delay}). 

\subsection{$|\hh|=1$}\label{One user in H}
In this case, there are two leaders, the user in $\hh$, $U_1\equiv U_h$ and $U_2 \equiv U_l$. For $d_{u_i}$ we have
\begin{equation}
\label{eq:du1}
d_{u_i}=\begin{cases}
\beta c_1=\lambda,& i=1 \quad (U_1\equiv U_h),\\
(\beta D^h_l+\bar{\beta})c_l,&i=2 \quad (U_2 \equiv U_l),\\
(\beta D_i^h+\bar{\beta} D_i^l)c_i,&i>2.
\end{cases}
\end{equation}

According to the delivery rate analysis in Section \ref{sec:Delivery Rate Analysis}, $\beta$ is the fraction of time that $\nn{1}$ is in the transmission queue, and $\uu{1}$ receives it with the probability of $c_1$. 
 Since all the packets received by $U_1$ are delivered, $d_{u_1}$ is given by (\ref{eq:du1}). On the other hand, the portion of time that  $U_l$ is the leader is  given by $\bar{\beta}$, and this user delivers the packets via leader transmissions with the probability of $\bar{\beta}c_l$ and differential knowledge transmissions  with the probability of $\beta D_l^hc_l$. Moreover, other users deliver the packets via differential knowledge transmissions from these two leaders with the probability given in (\ref{eq:du1}). Again, the delay probability is given by 
(\ref{eq:delivery delay}).
 
\subsection{ $|\hh|>1$} \label{More than one user in H}
When there are more than one user in $\hh$, all members of $\hh$ have a chance to be the leader and they receive differential knowledges from each other. Furthermore, using our analytical model, we cannot calculate the probability of being the leader and differential knowledge for the users in $\hh$. However, if we had the leader and  differential knowledge probabilities in $\hh$, then $d_{u_i}$ would be given by
\begin{equation}
\label{eq:2inH}
d_{u_i}=\begin{cases}
\displaystyle{\sum_{\substack{k:U_k \in \hh}}{\beta_kD_i^k}}c_i,&U_i\in \hh,\\
(\beta_hD_i^h+\bar{\beta}_hD_i^l)c_i,&U_i\in\LL.
\end{cases}
\end{equation}

 Where $\beta_k$ is the probability of $U_k\in \hh$ being the leader and $D_i^k$ is the probability of differential knowledge for $U_i$ when $U_k$ is the leader (Note that $D_i^i=1$ which corresponds to the leader transmission for $U_i$). Because we cannot calculate the values of  $\beta_k$s and  $D_i^k$s for the users in $\hh$, to evaluate the accuracy of (\ref{eq:2inH}), we extract these values from simulations and after calculating $d_{u_i}$, we use them in (\ref{eq:delivery delay}) and compare the results with simulations in Section \ref{sec:Simulation Results}
 (see Fig. \ref{pic:delay5_2}).
 
 \subsection{Expected value of the delay}\label{Expected value of the delay}
 Another parameter for comparing the delay of the users is the expected value of the delivery delay. 
 From (\ref{eq:delivery delay}), we have
 \begin{equation}
 \label{eq:delay expect}
 \mathbb{E}\{\theta_i\}=\sum_{T=0}^{\infty}{T\Pr(\theta_i=T)}=\frac{1}{R_i}.
 \end{equation} 
 This is a reasonable result that the average delay of each user has an inverse relation to its delivery rate.

%% file: 5_Simulation_Results.tex
\section{Simulation Results}\label{sec:Simulation Results}
In this section, we compare the  simulation results  with the analysis to validate the results given in the previous sections. In our simulation setup, we have considered different  packet arrival rates at the sender, number of users and channel erasure probabilities. These different settings are shown in Table \ref{tab:simconditions}. 

As described in Algorithm \ref{alg:codscheme}, at each time slot  a Gaussian elimination is performed to construct the transmitted packet $s(t)$. After each transmission, Gaussian elimination is again performed on the users' buffer to decode the packets. Then, the next required packet of each user (which is the oldest unseen packet) is determined. The newer seen packets are also stored in the buffer of each user until they receive older packets to decode them. Furthermore, for the purpose of the delivery delay analysis, the critical time slots for each packet such as arrival at the sender, seeing, decoding and delivering by each user are traced. Finally, we have measured the delivery rate and the delivery delay of the users after the delivery of $10000$ packets  to $\uu{1}$. 

\subsection{Delivery rate}
\begin{figure}[h]
	\centering
	\includegraphics[width=\columnwidth,height=0.65\columnwidth]{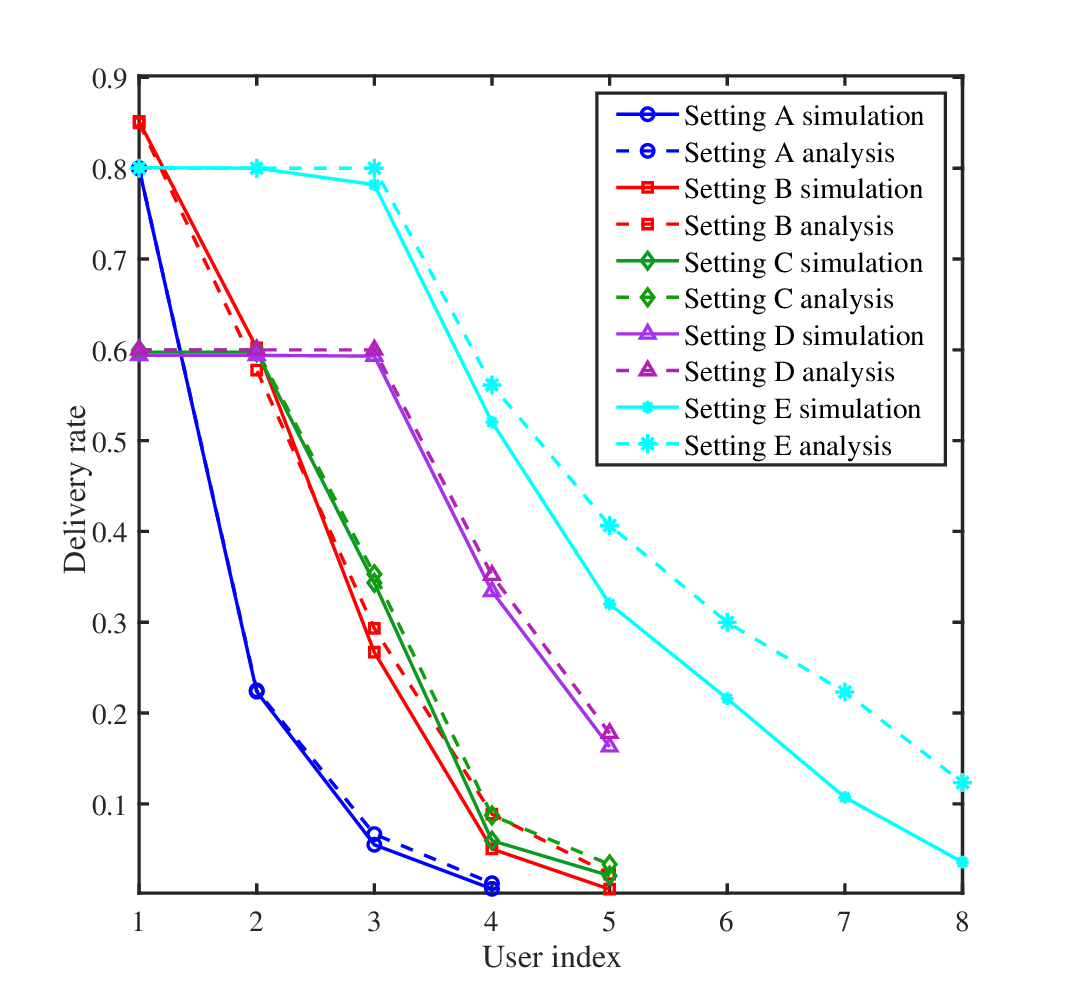}
	\caption{Delivery rate analysis and simulation for the settings of Table \ref{tab:simconditions}.  In analysis of the delivery rate, $\beta$ is given by (\ref{eq:beta}), differential knowledge is given by (\ref{eq:DH}) and (\ref{eq:DL}) and the delivery rate is given by (\ref{eq:Ri}).}
	\label{pic:delivery}
\end{figure}

The comparison of simulation and analysis for the delivery rates are depicted in Fig. \ref{pic:delivery}. 
To analyze delivery rate,  (\ref{eq:beta}) is used for the leader probability and  (\ref{eq:LH*})-(\ref{eq:LH}) have been used for the probability of leader transmissions, then  the differential knowledge probabilities are given by (\ref{eq:DH}), (\ref{eq:DL}) and the delivery rate is given by (\ref{eq:Ri}). 
In all settings, it is observed that the users in group $\hh$ have reached a delivery rate  very close to  $\lambda$, and for the users in $\mathcal{L}$,  there is a reasonable match between the simulation and analytical results. 

Among simulation settings, it can be observed that settings A and D have the minimum error margin. In setting A, since $\hh=\O$ and there is only one leader, there is no need for approximation and the simulation result is very close to analysis. In setting D, the error decreases due to the low value of $\lambda$ and the number of the users in $\LL$.  In setting C, $\lambda$ has the same value as D, however, there is one more user in $\LL$ and it causes the error to increase. In setting B, since there is only one user in $\hh$, the probability of being the leader is determined more accurately and
the error observed for the last users is due to the high value of $\lambda$. Except setting E, the pattern of the analyzed delivery rate is very similar to the simulations. These observations show that the accuracy of our model decreases when the number of users and the value of $\lambda$ increase.
\begin{table}[h]
	\caption{Simulation settings}
	\label{tab:simconditions}
	\centering
	\begin{tabular}{|c|c|c|}
		\hline Setting& $\lambda$ & $\boldsymbol{c}$ \\
		\hline A & $0.85$ & $[0.8,0.6,0.4,0.2]$ \\
		\hline B & $0.85$ & $[0.9,0.8,0.7,0.5,0.3]$ \\
		\hline C & $0.6$ & $[0.8, 0.7,0.5,0.3,0.2]$ \\
		\hline D & $0.6$ & $[0.9,0.8,0.7,0.5,0.4]$ \\
		\hline E & $0.8$ & $[0.9,0.85,0.8,0.75,0.7,0.65,0.6,0.5]$ \\
		\hline
	\end{tabular}
\end{table}

To have more accurate characterization of the delivery rates, as the number of users increases, more accurate analysis of leadership probability and differential knowledge is required. 
On the other hand, a key part of the analytical framework is Approximation \ref{app-RH}, which treats the leaders in group $\hh$ as a single user,  subsequently it affects the precision of analysis. 
In summary, to have a more accurate model, finding a way to determine the leader probability of the users in $\hh$  seems necessary.
\begin{figure}[]
	\centering
	\includegraphics[width=\columnwidth,height=0.65\columnwidth]{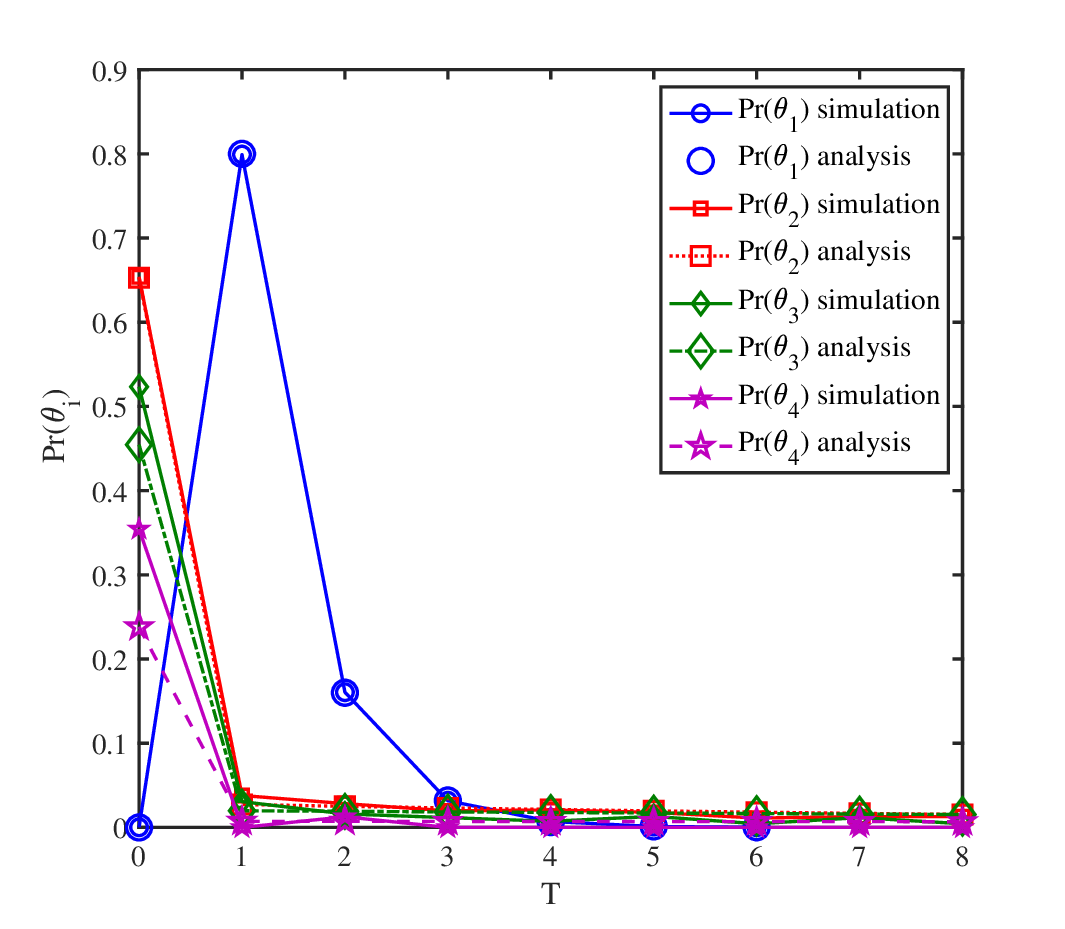}
	\caption{Simulation and analytical results for the probability of having $T$ time slots delivery delay ($\Pr(\theta_i=T)$) of Setting A in Table \ref{tab:simconditions}. To analyze the delivery delay, (\ref{NO_R_delay}) is used for $d_{u_i}$ and the delivery delay is given by (\ref{eq:delivery delay}).}
	\label{pic:delay4_0}
\end{figure}
\begin{figure}[]
	\includegraphics[width=\columnwidth,height=0.65\columnwidth]{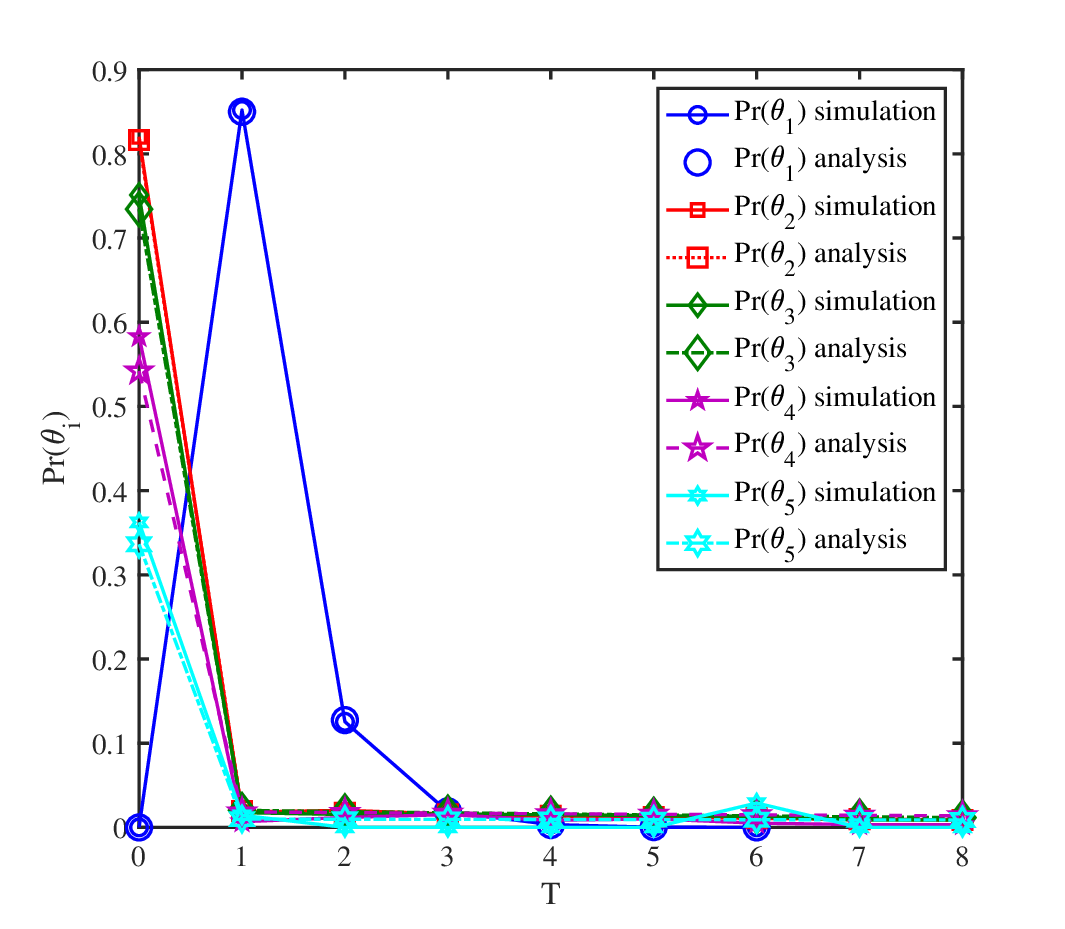}
	\caption{Simulation and analytical results for the probability of having $T$ time slots delivery delay ($\Pr(\theta_i=T)$) of Setting B in Table \ref{tab:simconditions}. To analyze the delivery delay, (\ref{eq:du1}) is used for $d_{u_i}$ and the delivery delay is given by (\ref{eq:delivery delay}).}
	\label{pic:delay5_1}
\end{figure}
\begin{figure}[h]
	\centering
	\includegraphics[width=\columnwidth,height=0.65\columnwidth]{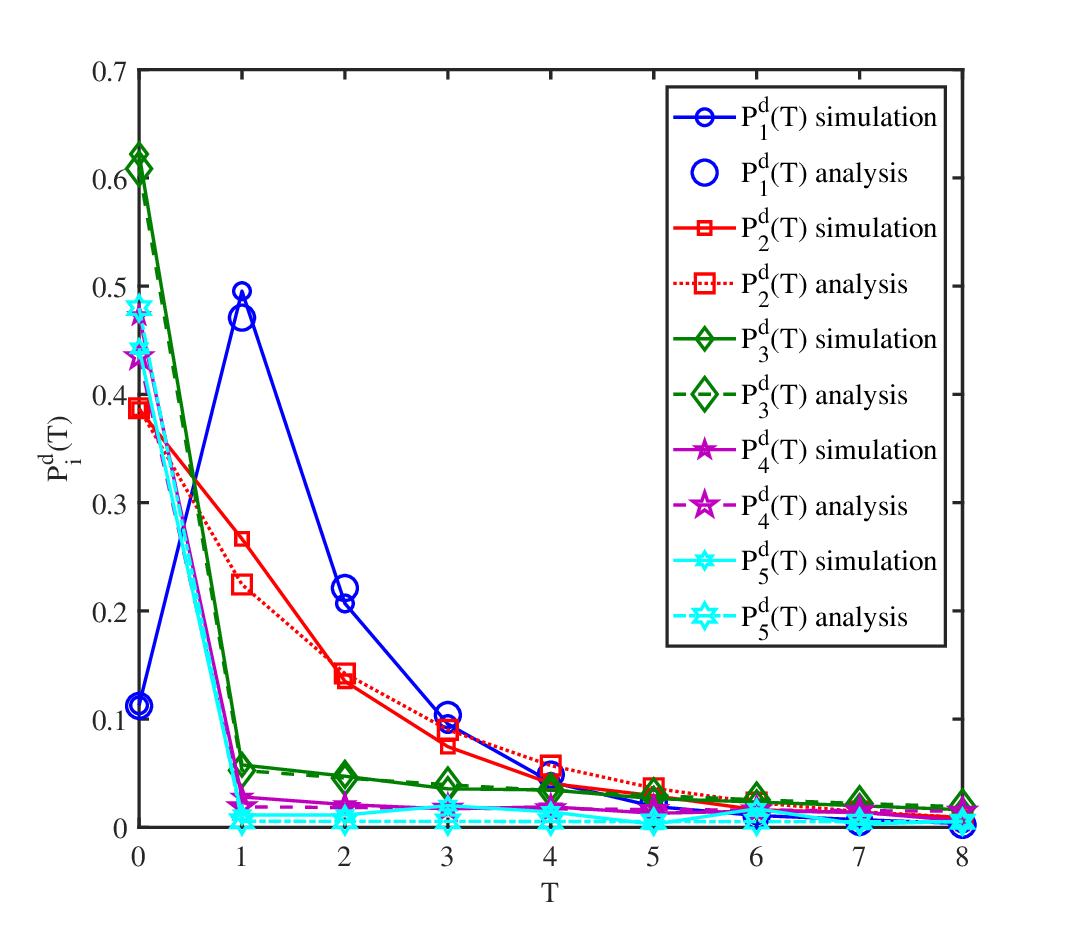}
	\caption{Simulation and analytical results for the probability of having $T$ time slots delivery delay ($\Pr(\theta_i=T)$) of Setting C in Table \ref{tab:simconditions}. To analyze the delivery delay, (\ref{eq:2inH}) is used for $d_{u_i}$ and the delivery delay is given by (\ref{eq:delivery delay}).}
	\label{pic:delay5_2}
\end{figure}

\begin{figure}[t]
	\includegraphics[width=\columnwidth,height=0.65\columnwidth]{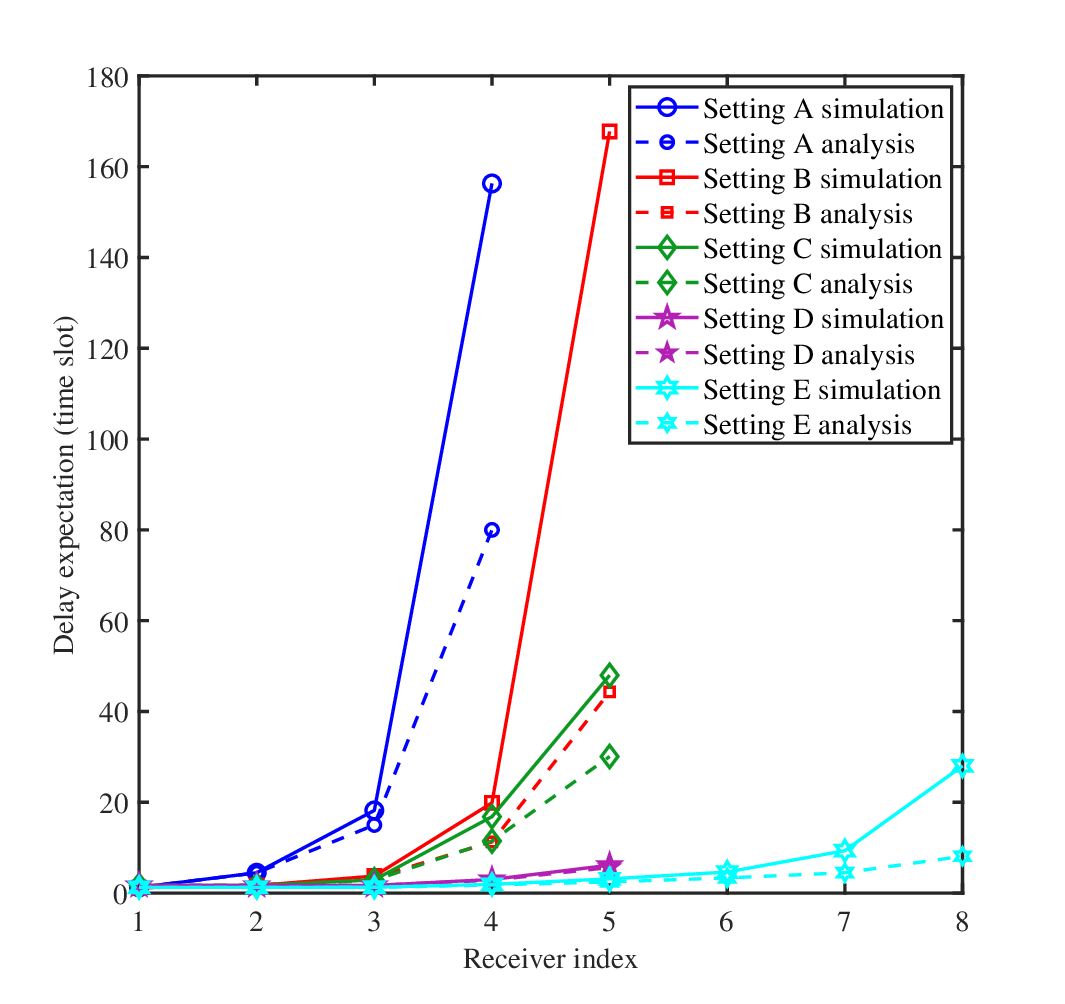}
	\caption{Delay expectation of different users, analysis and simulation for the settings of Table \ref{tab:simconditions}. For analysis (\ref{eq:delay expect}) is used.}
	\label{pic:delay expect}
\end{figure}

\subsection{Delivery Delay}
Here, the derived expressions for the delay probabilities are compared with the values of simulations. Fig. \ref{pic:delay4_0} and Fig. \ref{pic:delay5_1} illustrate the simulation and analysis results for the cases of Section \ref{No user in group H} and \ref{One user in H} respectively. For these cases, the settings A and B of Table \ref{tab:simconditions} have been used for the simulation and (\ref{eq:delivery delay}), (\ref{NO_R_delay})  and (\ref{eq:du1})  have been used for the analysis. As it is observed, the analysis shows perfect match with simulation for $U_1$ and $U_2$ and loses its accuracy for the other users in the both settings due to the error in the calculation of  differential knowledge probability. It is noteworthy that $\Pr(\theta_1=0)$ is zero, that shows $U_1$ did not receive non-differential packets, because in setting A, $U_1$ has been the leader in all time slots, and in setting B,  $U_2$ could be the leader only when  $N_1(t)$ is not in the transmission queue. Moreover, for $U_1$ in the both settings most of the packets are delivered with the delay of $T=1$ i.e., in the next time slot after requesting the packets. However, for the other users $\Pr(\theta_i=0)$ is maximum, since they receive  most of their packets by the non-differential transmissions from the leaders. Although $\Pr(\theta_i=0)$ is maximum, it does not mean that the users experience low delay. In order to compare the users in terms of the delay they experience, we should look at the range of $T$ for each user. For instance, in setting A, $U_1$ has the range of $0\leq T\leq 6$ while for $U_2$ the range is $0 \leq T \leq 90$ (the ranges in the figures are limited to have better illustration). The maximum value of $T$ increases for $U_3$ to $202$ and for the last user to $1100$. Furthermore, for $U_i$'s with $i>1$ the probability of delay has a slow decline for $T>0$ that shows the number of packets for each value of $T$ is close to each other.

The simulation results for Section \ref{More than one user in H}  is depicted in Fig. \ref{pic:delay5_2}. Setting C of Table \ref{tab:simconditions}, (\ref{eq:delivery delay}) and (\ref{eq:2inH}) have been used for the simulation and analysis. Note that in this case  $\displaystyle\Pr(\theta_1=0)$ is not zero, because the other members in $\hh$ could be the leader and all of the users in $\hh$ can receive differential knowledge and also non-differential transmissions from each other.
Figure \ref{pic:delay expect} illustrates the simulation and analysis results for \ref{eq:delay expect}.  Since expectation of the delay is the inverse of the delivery rate, error margin of  the delay expectation increases for weaker users because the delivery rate value of these users is small and a little error in its analysis  affects the delay expectation considerably. Using this comparison, we conclude that the stronger users have less delay.

%% file: 6_Conclusions.tex
\section{Conclusions} \label{sec:Conclusions}
In this paper, we have shown that a previously proposed network coding scheme can achieve efficient multirate packet delivery in heterogeneous broadcast packet erasure networks.  Also, we have introduced an appropriate model to  estimate the delivery rate and the delivery delay of the system.  Using this coding scheme,  the strongest user, receives packets with the maximum possible throughput and the other users have a non-zero delivery rate according to their link capacities.
Moreover, we have introduced a new  definition for the delivery delay and analyzed the system based on it. The number of time slots between the first request of a packet and its delivery is counted as the delay. Using this definition, the delivery delays of the users have been compared and a simple expression has been derived. Similar to the delivery rate analysis, the numerical results for the delivery delay have shown a reasonable match between our analysis and the simulation results, especially for stronger users. 
Although achieving multirate packet delivery is possible for a number of users, it seems by increasing the number of  users, the delivery rate of the weaker users tends to zero.  Designing a coding method to support multirate packet delivery for a large number of users in  a heterogeneous case can be considered in future works. Furthermore,  considering other performance measures like fairness might also be useful.